\documentclass[]{fairmeta}
\usepackage{amsfonts}
\usepackage{url}
\usepackage{xspace}
\usepackage{graphicx}
\usepackage{multirow} 
\usepackage{booktabs}
\usepackage{algorithm}
\usepackage{algorithmic}
\usepackage{wrapfig}
\usepackage{enumitem}
\usepackage{rotating}
\usepackage{siunitx}
\usepackage[most]{tcolorbox}
\usepackage{float}
\usepackage[table]{xcolor}
\usepackage{wrapfig}

\newcommand{\MODEL}{\textsc{MetaEmbed}\xspace}

\newcommand{\EMBED}{\emph{Meta Embeddings}\xspace}
\newcommand{\TOKEN}{\emph{Meta Tokens}\xspace}

\makeatletter
\DeclareRobustCommand\onedot{\futurelet\@let@token\@onedot}
\def\@onedot{\ifx\@let@token.\else.\null\fi\xspace}

\def\ie{\emph{i.e}\onedot}

\makeatother

\newcommand{\custompara}[1]{{\vspace{1mm}\noindent\textbf{#1}\xspace}}
\newcommand{\newpara}[1]{\textbf{#1}\xspace}

\title{MetaEmbed: Scaling Multimodal Retrieval at Test-Time with Flexible Late Interaction}

\author[1,2,*]{Zilin Xiao}
\author[1]{Qi Ma}
\author[1]{Mengting Gu}
\author[1]{Jason Chen}
\author[1]{Xintao Chen}
\author[2]{Vicente Ordonez}
\author[1]{Vijai Mohan}

\affiliation[1]{Meta Superintelligence Labs}
\affiliation[2]{Rice University}

\contribution[*]{Work done at Meta}

\abstract{
Universal multimodal embedding models have achieved great success in capturing semantic relevance between queries and candidates. 
However, current methods either condense queries and candidates into a single vector, potentially limiting the expressiveness for fine-grained information, or produce too many vectors that are prohibitive for multi-vector retrieval.
In this work, we introduce \MODEL, a new framework for multimodal retrieval that rethinks how multimodal embeddings are constructed and interacted with at scale.
During training, a fixed number of learnable Meta Tokens are appended to the input sequence.
At test-time, their last-layer contextualized representations serve as compact yet expressive multi-vector embeddings. 
Through the proposed Matryoshka Multi-Vector Retrieval training, \MODEL learns to organize information by granularity across multiple vectors.
As a result, we enable test-time scaling in multimodal retrieval where users can balance retrieval quality against efficiency demands by selecting the number of tokens used for indexing and retrieval interactions. 
Extensive evaluations on the Massive Multimodal Embedding Benchmark (MMEB) and the Visual Document Retrieval Benchmark (ViDoRe) confirm that \MODEL achieves state-of-the-art retrieval performance while scaling robustly to models with 32B parameters.
}

\date{\today}
\metadata[Code]{\url{https://github.com/facebookresearch/MetaEmbed}}
\correspondence{Zilin Xiao at \email{zilin@meta.com}}

\begin{document}

\maketitle

\section{Introduction}
Multimodal embedding models play an essential role in image search~\citep{gordo2016deep}, visual question answering~\citep{hu2018learning, zheng2021knowledge} and visual document retrieval~\citep{ColPali}, where models project heterogeneous inputs into a unified representation space. 
While existing methods, including CLIP~\citep{CLIP}, BLIP~\citep{BLIP} and SigLIP~\citep{SigLIP} have demonstrated superior performance in cross-modal retrieval, their performance remains limited in scenarios where the inputs involve complex and diverse instructions. 
Thanks to recent advances in building embeddings through foundation vision-language models (VLMs), one could apply contrastive learning on the extracted embedding from the hidden states of the last layer of a VLM to learn meaningful multimodal representations while retaining pre-trained knowledge. 

Despite growing progress in multimodal embedding VLMs, the common practice of condensing the entire query and candidate into a single vector is not an optimal choice, as fine-grained details are lost between modalities~\citep{yao2022filip, thrush2022winoground} and this process has theoretical limitations~\citep{weller2025theoretical}. 
In text retrieval, ColBERT~\citep{khattab2020colbert} introduced a multi-vector late interaction mechanism that retains multiple token-level embeddings and uses a lightweight scoring between query and document token representations.
This approach preserves significantly more contextual information than single-vector methods while still allowing independent encoding of queries and documents, and has motivated a recent trend of devising multi-vector embeddings for multimodal retrieval~\citep{ColPali, xu2025llama, jinaembedding}. 

However, multi-vector methods incur substantial efficiency costs in terms of \textit{index size}, \textit{retrieval latency} and \textit{feasibility}. 
In these methods, each image is encoded into hundreds of patch embeddings, and each query text into several token embeddings, all of which must be stored and compared during retrieval. 
This results in large index sizes and slower retrieval processing.
Moreover, multimodal-to-multimodal retrieval becomes impractical when both the query and candidate sides contain images, as similarity computation for each query-candidate pair involves interactions between thousands of query tokens and thousands of candidate tokens, making both training and inference prohibitive due to computational demands. 

\begin{figure*}[t]
	\begin{center}
		\includegraphics[width=0.95\linewidth]{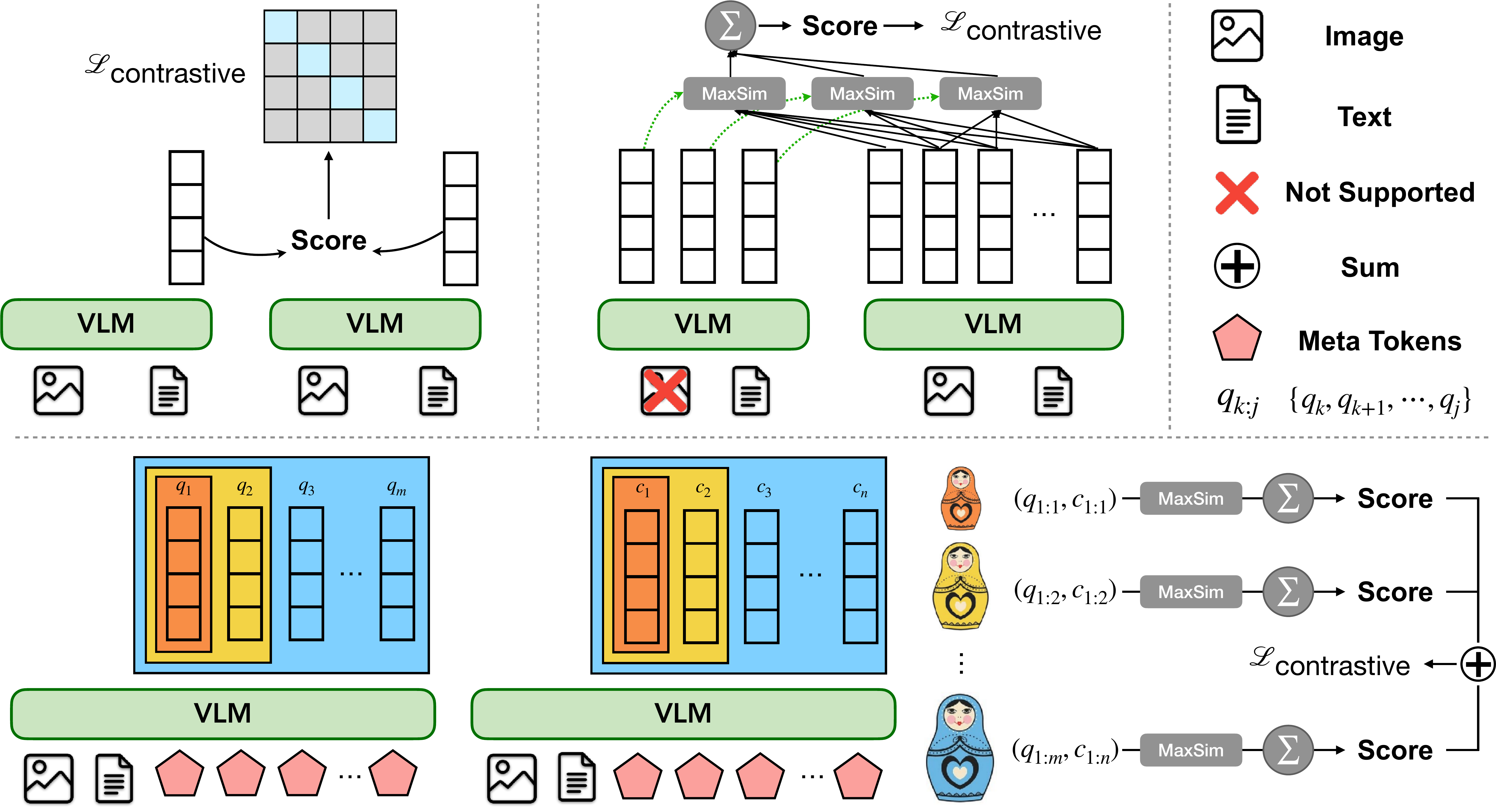}
	\end{center}
	\caption{
    {\bf Upper Left: }Single vector retrieval method computes a score for each pair of query and candidate and uses a contrastive objective to maximize the score for corresponding pairs. 
    {\bf Upper Right: }Multi-vector retrieval aggregates maximum similarities across vector pairs before training. 
    {\bf Lower: } \MODEL structures query and candidate vectors into hierarchical nested groups and trains coarse-to-fine multi-vector embeddings that enable scalable and flexible retrieval.
}
	\label{fig:teaser}
\end{figure*}

In this work, we propose \MODEL as a scalable late-interaction training recipe that advances multimodal retrieval with a flexible multi-vector method, illustrated in Figure~\ref{fig:teaser}.
Instead of encoding the query and candidate into one vector, we introduce a small number of learnable \TOKEN appended to the input sequence of the query and candidate. Their last-layer hidden states serve as a set of contextualized representations for late interaction, namely \EMBED.
To enable flexible late interaction, where users can trade off retrieval accuracy against computational budget and retrieval latency, we draw inspiration from Matryoshka Representation Learning~\citep{kusupati2022matryoshka} and design the Matryoshka Multi-Vector Retrieval (MMR) module in \MODEL. 
By performing contrastive learning across parallel nested groups of representations \textbf{at training-time}, the model learns coarse-to-fine multi-vector embeddings that can be selectively utilized for late interactions depending on the computation budget \textbf{at test-time}. 
Increasing the number of \EMBED used at indexing improves the retrieval quality at the cost of index storage budget and retrieval latency, thus enabling test-time scaling in multimodal retrieval.

We first validate \MODEL on the Massive Multimodal Embedding Benchmark (MMEB)~\citep{MMEB} and Visual Document Retrieval Benchmarks (ViDoRe) v2~\citep{ColPali, mace2025vidore}, which represent a comprehensive suite of retrieval tests covering images, text and visual documents. 
Our experiments show that \MODEL achieves state-of-the-art retrieval performance across diverse scenarios. 
To further examine its generality and training scalability, we evaluate \MODEL with different VLM architectures and model sizes.
Notably, test-time scalability remains effective even at 32B scale, with minimal diminishing returns as models grow larger. 
We hope \MODEL charts a path toward multimodal retrieval systems that are both accurate and deployable at scale, advancing the pursuit of generality, efficiency, and flexibility.

\section{Related Work}
\custompara{Multimodal Embedding.} 
These methods aim to project heterogeneous inputs into a shared representation space for cross-modal understanding and retrieval (e.g~\cite{devise,kiros2014unifying,faghri2018vse++}).
More recent large scale models such as CLIP~\citep{CLIP}, MetaCLIP~\citep{xu2023demystifying, chuang2025meta}, BLIP~\citep{BLIP} and SigLIP~\citep{SigLIP} encode each modality independently and apply contrastive training to enforce cross-modal alignment. 
More recent methods are built upon stronger VLMs~\citep{xiao2024grounding, kong2025modality, qin2025unimoco, ju2025generator, lin2025mmembed, xiao2025locore}. For instance, VLM2Vec~\citep{jiang2025vlmvec} adapts Phi-3.5-V~\citep{abdin2024phi3technicalreporthighly}, VLM2Vec-V2~\citep{meng2025vlm2vec} and GME~\citep{zhang2024gme} builds on Qwen2~\citep{Qwen2-VL} and LLaVE~\citep{lan2025llave} finetunes on LLaVA~\citep{li2024llava}. 
Beyond architectural choices, the community has also explored innovative strategies in data construction and training. 
MegaPairs~\citep{megapairs} and mmE5~\citep{mme5} curate large-scale synthetic data with sophisticated pipelines to support contrastive training. 
UniME~\citep{gu2025breaking} achieves strong results through diverse data combined with teacher model distillation. 
B3~\citep{B3} incorporates novel insights into batch mining techniques. 
MoCa~\citep{chen2025moca} used continual pre-training to produce bidirectional embeddings.
Nevertheless, many existing multimodal retrieval methods predominantly rely on single-vector retrieval, which hinders further scaling as embedding size becomes a bottleneck.

\custompara{Multi-Vector Retrieval.}
Multi-vector retrieval refers to a family of dense retrieval methods that represent queries and documents with multiple embeddings rather than a single vector~\citep{toliasiclr2016, NEURIPS2019_471c75ee, BMVC2017_89}, with ColBERT~\citep{khattab2020colbert} being a successful recent example of this paradigm by introducing a late interaction framework. 
While many variants have been proposed to improve multi-vector retrieval efficiency through approximation~\citep{lee2023rethinking, engels2023dessert, jayaram2024muvera} and compression~\citep{santhanam2022colbertv2, santhanam2022plaid, li-etal-2023-citadel}, naive ColBERT-style methods such as ColPali, ColQwen~\citep{ColPali} and others~\citep{jinaembedding, xu2025llama} still remain dominant in the context of text-image retrieval. 
However, these approaches do not support multimodal-to-multimodal retrieval, since introducing hundreds of image tokens on the query side renders both training and inference computationally prohibitive, highlighting the need for our proposed \MODEL framework. 

\custompara{Matryoshka Representation Learning.}
Matryoshka Representation Learning (MRL)~\citep{kusupati2022matryoshka} was introduced to encode features at multiple granularities within a single vector in a nested structure. 
Popular text-only single-vector retrieval models~\citep{zhang2025qwen3, jinaembedding} natively support MRL, enabling retrieval to dynamically select the first few dimensions according to the available computational budget.
Although prior work~\citep{cai2025matryoshka} has applied Matryoshka methods for token budgeting in VLM generation, to the best of our knowledge, \MODEL presents the first work that leverages such a framework for multi-vector retrieval and achieves successful test-time scaling. 
\section{Methodology}
In this section, we first revisit the definition of multimodal retrieval and how late interaction works to utilize multiple vectors for retrieval. 
Then we introduce the \MODEL recipe, its model architecture, and how it enables test-time scaling in multimodal retrieval. 

\subsection{Preliminaries}

\custompara{Problem Definition.} 
Multimodal retrieval consists of retrieving relevant content across different modalities, where the query $\mathbf{q}$ can be text $q_t$, an image $q_i$ or a combination of both $(q_t, q_i)$. And the retrieval candidates $\mathbf{c}$ can likewise be of any modality or multimodal combination. 
Given a query $\mathbf{q}$ and a set of $\mathbf{N}$ candidates $\{\mathbf{c}_1, \mathbf{c}_2, \ldots, \mathbf{c}_N\}$, a multimodal retrieval model typically defines a similarity function $s(\mathbf{q}, \mathbf{c})$ to measure the relevance between $\mathbf{q}$ and a candidate $\mathbf{c}$. The retrieved top-1 prediction $\mathbf{c}^\star$ is then determined by:
\begin{equation}
    \mathbf{c}^\star = \underset{\mathbf{c} \in \{\mathbf{c}_1, \ldots, \mathbf{c}_N\}}{\arg\max} \; s(\mathbf{q}, \mathbf{c}).
\end{equation}

\custompara{Late Interaction.}
For a query $\mathbf{q}$ and a candidate $\mathbf{c}$, let their multi-vector representations be denoted as $\mathbf{E}_{\mathbf{q}} \in \mathbb{R}^{N_q \times D}$ and $\mathbf{E}_{\mathbf{d}} \in \mathbb{R}^{N_d \times D}$, where $D$ is the embedding dimension, and $N_q, N_d$ are the number of token-level vectors for the query and the candidate, respectively. The late interaction operator $\mathbf{LI}(q,d)$ captures the most informative alignment by selecting, for each query vector $\mathbf{E}_{\mathbf{q}}^{(i)}$, its maximum similarity (dot product) with the document vectors $\mathbf{E}_{\mathbf{d}}^{(j)}$, and summing across all query vectors:
\begin{equation}
    \label{eq:late_interaction}
    \mathbf{LI}(q, d) = \sum_{i=1}^{N_q} \max_{j \in [1, N_d]} \left\langle \mathbf{E}_{\mathbf{q}}^{(i)}, \mathbf{E}_{\mathbf{d}}^{(j)} \right\rangle.
\end{equation}

\subsection{Our Design}

\custompara{\MODEL Recipe.}
\MODEL is designed as a scalable late-interaction retrieval model that introduces a small number of learnable \TOKEN appended to the input sequence of both queries and candidates. 
These \TOKEN are processed jointly with the original input by an underlying Vision-Language Model (VLM), and their final hidden states serve as \EMBED. 
Unlike patch- or token-level embeddings, \EMBED provide a set of compressed yet expressive vectors that capture fine-grained semantics through contextualization. 
This design drastically reduces the number of vectors required for retrieval while maintaining retrieval quality. 

Formally, let a VLM with parameters $\boldsymbol{\theta}$ define a conditional probability distribution 
$p_{\boldsymbol{\theta}} \left( \mathbf{y} \mid \mathbf{x}, \mathcal{I} \right)$
where $\mathbf{x}=[x_1,\dots,x_n]$ is the query or document text prompt and $\mathcal{I}$ are associated input images. 
\MODEL augments the input with learnable \TOKEN: queries use $\mathbf{M}_q\in\mathbb{R}^{R_q\times D}$, and candidates use $\mathbf{M}_c\in\mathbb{R}^{R_c\times D}$. For an input $(\mathbf{x},\mathcal{I})$, the transformer consumes  
$\mathbf{z}^{(0)} = \big[\; \mathbf{v}\ ;\ \mathbf{t}\ ;\ \mathbf{M}_q\ ;\ \mathbf{M}_c  \;\big]\in\mathbb{R}^{(P+n+R_q+R_c)\times D},$
where $\mathbf{v}$ and $\mathbf{t}$ are $P$ visual patches tokens and text inputs. 
The model produces last-layer hidden states $\mathbf{H}=F_{\boldsymbol{\theta}}(\mathbf{z}^{(0)})\in\mathbb{R}^{(P+n+R_q+R_c)\times D}$, where $F_{\boldsymbol{\theta}}$ denotes the transformer network parameterized by $\boldsymbol{\theta}$.
We extract the final hidden states at the \TOKEN positions to obtain query-side embeddings $\mathbf{E}_{\text{meta}}^{(q)} \in \mathbb{R}^{R_q\times D}$ or candidate-side embeddings $\mathbf{E}_{\text{meta}}^{(c)} \in \mathbb{R}^{R_c\times D}$, followed by L2 normalization. Each $\mathbf{E}_{\text{meta}}^{(q)}$ and $\mathbf{E}_{\text{meta}}^{(c)}$ constitutes a compact, contextualized \emph{multi-vector} representation produced in two separate forward passes of the VLM. 

\custompara{Matryoshka Multi-Vector Retrieval (MMR). }
With $\mathbf{E}_{\text{meta}}^{(q)}\in\mathbb{R}^{R_q\times D}$ and $\mathbf{E}_{\text{meta}}^{(c)}\in\mathbb{R}^{R_c\times D}$ available, we can compute a late-interaction score between a query $\mathbf{q}$ and a candidate $\mathbf{c}$ as follows:  
\begin{equation}
s(\mathbf{q},\mathbf{c})=\sum_{i=1}^{R_q}\max_{j\in[1,R_c]}\left\langle \mathbf{E}_{\mathbf{q}}^{(i)}, \mathbf{E}_{\mathbf{c}}^{(j)}\right\rangle .
\end{equation}
While effective, using all vectors for every instance is not flexible: the index size scales as $O(N \times R_c \times D)$ for $N$ candidates, and the scoring cost scales as $O(R_q \times R_c \times D)$ per pair.  
We therefore seek a mechanism that maintains strong retrieval quality under tight resources and scales to higher accuracy as more compute is allocated.
Inspired by~\cite{kusupati2022matryoshka}
, we impose a \emph{prefix-nested} structure on \EMBED so that the first few vectors form a coarse summary, and additional vectors refine the representation.  Concretely, fix $G$ group sizes for queries so that
\[
1 \le r_q^{(1)} < r_q^{(2)} < \cdots < r_q^{(G)} = R_q,
\]  
and for candidates so that
\[
1 \le r_c^{(1)} < r_c^{(2)} < \cdots < r_c^{(G)} = R_c.
\]  
For any input, define the $g$-th group of query embeddings as $\mathbf{E}^{(q,g)}=\mathbf{E}_{\text{meta}}^{(q)}[1{:}r_q^{(g)},:]$, and similarly for candidates $\mathbf{E}^{(c,g)}=\mathbf{E}_{\text{meta}}^{(c)}[1{:}r_c^{(g)},:]$. We then compute group-specific late-interaction scores  
\begin{equation}
s^{(g)}(\mathbf{q},\mathbf{c})=\sum_{i=1}^{r_q^{(g)}}\max_{j\in[1,r_c^{(g)}]}\left\langle \mathbf{E}_{\mathbf{q}}^{(g,i)}, \mathbf{E}_{\mathbf{c}}^{(g,j)}\right\rangle .
\end{equation}
During training, we optimize contrastive objectives across all groups in parallel, encouraging each prefix to be discriminative on its own while remaining consistent with larger prefixes.

\custompara{Training Objective.} 
Let $\mathcal{B}={(\mathbf{q}^{(b)},\mathbf{c}^{(b)},\mathbf{c}^{(b,-)})}_{b=1}^{B}$ be a minibatch where each query has a corresponding positive candidate $\mathbf{c}^{(b)}$ and one additional hard negative $\mathbf{c}^{(b,-)}$. For each group $g$, we define the similarity scores between query $u$ and candidate $v$ as follows:  
\begin{equation}     
\mathbf{S}^{(g)}_{u,v}=\tfrac{1}{\tau}\,s^{(g)}\!\big(\mathbf{q}^{(u)},\mathbf{c}^{(v)}\big),
\end{equation}  
with $\tau>0$ as a temperature hyper-parameter. For query $u$, the denominator of the softmax spans (i) all in-batch candidates $\{\mathbf{c}^{(1)},\ldots,\mathbf{c}^{(B)}\}$ and (ii) its explicit hard negative $\mathbf{c}^{(u,-)}$. The InfoNCE loss~\citep{oord2018representation} for group $g$ is:  
\begin{equation}     
\label{eq:nce_loss}
\mathcal{L}^{(g)}_{\text{NCE}} =-\frac{1}{B}\sum_{u=1}^B \log \frac{\exp(\mathbf{S}^{(g)}_{u,u})} {\exp(\mathbf{S}^{(g)}_{u,u})+\sum_{v\neq u}\exp(\mathbf{S}^{(g)}_{u,v})+\exp(\tfrac{1}{\tau}s^{(g)}(\mathbf{q}^{(u)},\mathbf{c}^{(u,-)}))}.
\end{equation}  
The final loss combines all groups with group-specific hyper-parameters $w_g$ as importance scales:
\begin{equation}
    \label{eq:final_loss}
    \mathcal{L}_{\text{final}}=\sum_{g=1}^G w_g\,\mathcal{L}^{(g)}_{\text{NCE}}.
\end{equation}

\begin{figure*}[t]
	\begin{center}
		\includegraphics[width=\linewidth]{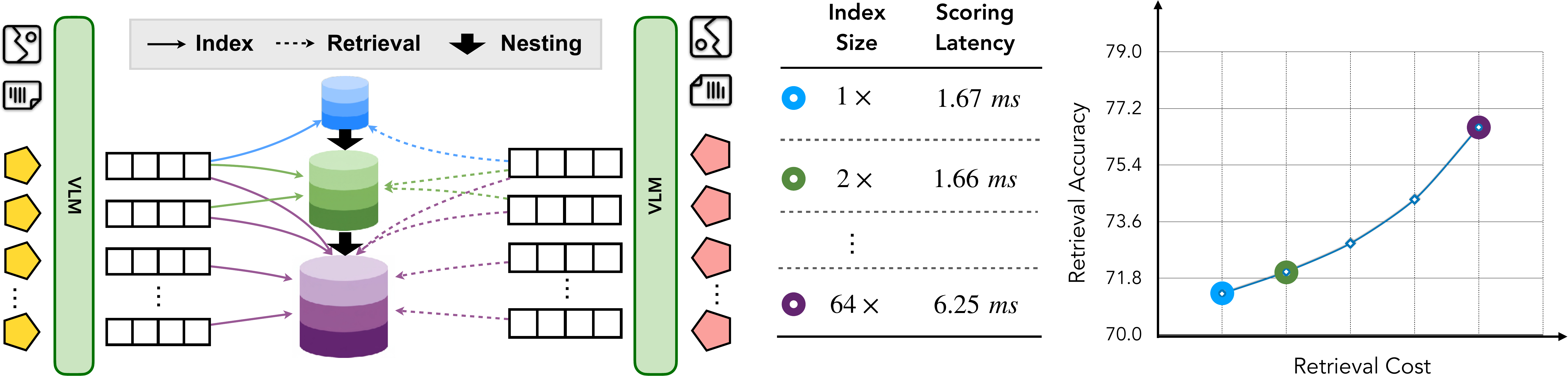}
	\end{center}
	\caption{Illustration of test-time scaling with varying retrieval budgets. 
    {\bf Left:} \MODEL constructs a nested multi-vector index that can be retrieved flexibly given different budgets. 
    {\bf Middle:} How the scoring latency grows with respect to the index size. Scoring latency is reported with 100,000 candidates per query on an A100 GPU.
    See \S\ref{sec:discussion} for full efficiency analysis. 
    {\bf Right: } \MODEL-7B performance curve with different retrieval budgets.
    See Figure~\ref{fig:testing_scale} (b) for full metrics.
    }
	\label{fig:scaling}
    \vspace{-0.2cm}
\end{figure*}

\custompara{Test-time Scaling.} 
The nested design yields a simple accuracy-efficiency knob, as illustrated in Figure~\ref{fig:scaling}. At indexing time, one may store only the first $r_c^{(g)}$ vectors for each candidate. At query time, the system selects $(r_q^{(g)}, r_c^{(g)})$ based on latency constraints and computes $s^{(g)}(\mathbf{q},\mathbf{c})$ for scoring. Coarse prefixes ($g$ small) are ideal for fast retrieval scoring, while larger prefixes ($g$ large) improve precision at the expense of additional compute. 
Because those groups are optimized in parallel, we can seamlessly adjust the retrieval granularity and budget without retraining the system by selecting a different group size at test-time.
In later sections, we refer to the selected combination of group sizes $(r_q^{(g)}, r_c^{(g)})$ as the \textbf{retrieval budget}.

\section{Experiments}
In this section, we first introduce experimental settings, including models, training data and benchmarks. We then report comprehensive results to showcase the effectiveness and robustness of \MODEL. 

\subsection{Settings}
\custompara{Models. }
To evaluate the effectiveness of \MODEL as a training recipe, we conduct experiments on various VLMs of different sizes, including Qwen2.5-VL~\citep{Qwen25-VL}, PaliGemma~\citep{beyer2024paligemma} and Llama-3.2-Vision~\citep{grattafiori2024llama}. 
Qwen2.5-VL and PaliGemma represent unified multimodal architectures that process text and vision inputs, while Llama-3.2-Vision represents cross-attention-based designs where visual information is integrated into the language model through cross-attention layers.
In this section, \MODEL-3B, -7B and -32B refer to models finetuned on Qwen2.5-VL backbones and \MODEL-11B is finetuned on the Llama-3.2-Vision model.

\custompara{Training. }
We train \MODEL-7B on 32 NVIDIA H100 SXM5 96GB GPUs for 3,500 steps with a global batch size of $2,048$, which leads to 30 hours for training. Appendix~\ref{app:imple} has training details for other variants.
We use LoRA~\citep{lora} with a rank of 32 and scaling factor $\alpha=32$ in all training. 
For models with Matryoshka Multi-Vector Retrieval, 
we empirically choose $G=5$ group sizes of $(r_q, r_c)$ as $\{(1,1), (2,4), (4,8), (8,16), (16,64)\}$ and discuss other group size options in \S\ref{sec:ablations}.
Group-specific hyper-parameter $w_g$ in Equation~\ref{eq:final_loss} is set to 1 following~\cite{kusupati2022matryoshka}.
Contrastive training temperature $\tau$ is set to $0.03$.
We only incorporate 
MMEB-train~\citep{jiang2025vlmvec} and ViDoRe-train~\citep{ColPali} with one explicit hard negative from \cite{chen2025moca} for training all variants of \MODEL.

\custompara{Evaluation. }
We assess the general multimodal embedding ability of \MODEL on the Massive Multimodal Embedding Benchmark (MMEB)~\citep{MMEB} and use Precision@1 as the evaluation metric. MMEB is an established benchmark covering 36 tasks across four types, including classification, visual question answering (VQA) e.g. ScienceQA~\citep{NEURIPS2022_11332b6b}, VizWiz~\citep{Gurari_2018_CVPR}, ChartQA~\citep{masry-etal-2022-chartqa}, retrieval across a variety of domains e.g. Visual News~\citep{liu-etal-2021-visual}, FashionIQ~\citep{Wu_2021_CVPR}, OvenWiki~\citep{Hu_2023_ICCV}, and visual grounding e.g. COCO~\citep{coco}, RefCOCO~\citep{referitgame,refcoco}. 
In addition, to compare with existing multi-vector solutions on text-image retrieval, we evaluate \MODEL on Visual Document Retrieval Benchmarks (ViDoRe) v2~\citep{mace2025vidore} and use average NDCG@5 as the metric. 
ViDoRe~\citep{ColPali} was first introduced to benchmark visual document retrieval capabilities in different domains, and its v2 version mitigates performance saturation by including more generalized settings and incorporating multilingual subsets. 
We refer to Appendix~\ref{app:baseline_method_intro} for an introduction to selected baseline methods.

\begin{table}[t]
\centering
\small
\setlength{\tabcolsep}{5pt}
\caption{Precision@1 (\%) results on MMEB, which includes 36 tasks across four categories: Classification, Visual Question Answering (VQA), Retrieval, and Visual Grounding. IND and OOD represent the in-domain average and out-of-domain average metrics, respectively. 
\textbf{Bold} denotes the best scores in the subset and the second-best scores are highlighted with \underline{underline}.}
\vspace{3pt}
\begin{tabular}{lcccccccc}
\toprule
\multirow{2}{*}{Models} & \multirow{2}{*}{Size} & \multicolumn{4}{c}{\bf Per Meta-Task Score} & \multicolumn{3}{c}{\bf Average Score} \\
\cmidrule(r){3-6}\cmidrule(l){7-9}
 &  & Classification & VQA & Retrieval & Grounding & IND & OOD & Overall \\
\midrule
\rowcolor{gray!15}
\multicolumn{9}{c}{Baseline Models} \\
CLIP                    & 428M   & 55.2 & 19.7 & 53.2 & 62.2 & 47.6 & 42.8 & 45.4 \\
MagicLens           & 613M   & 38.8 & 8.3  & 35.4 & 26.0 & --   & --   & 27.8 \\
UniIR                   & 428M   & 42.1 & 15.0 & 60.1 & 62.2 & --   & --   & 42.8 \\
ABC  &  7B  &  60.0  & 31.0  & -- & -- & -- & -- & -- \\
MM-EMBED             & 7B   & 48.1 & 32.2 & 63.8 & 57.8 & --   & --   & 50.0 \\
GME                      & 7B   & 56.9 & 41.2 & 67.8 & 53.4 & --   & --   & 55.8 \\
VLM2Vec                  & 7B   & 61.2 & 49.9 & 67.4 & 86.1 & 67.5 & 57.1 & 62.9 \\
VLM2Vec-V2               & 2B   & 62.9 & 56.3 & 69.5 & 77.3 & -- & -- & 64.9 \\
MMRet              & 7B   & 56.0 & 57.4 & 69.9 & 83.6 & 68.0 & 59.1 & 64.1 \\
mmE5                     & 11B   & 67.6 & 62.7 & 71.0 & 89.7 & 72.4 & 66.6 & 69.8 \\
MoCa-3B & 3B & 59.8 & 62.9 & 70.6 & 88.6 & 72.3 & 61.5 & 67.5 \\
MoCa-7B & 7B & 65.8 & 64.7 & 75.0 & \textbf{92.4} & 74.7 & 67.6 & 71.5 \\
B3-7B & 7B & 70.0 & 66.5 & 74.1 & 84.6 & 75.9 & 67.1 & 72.0 \\
\midrule
\rowcolor{gray!15}
\multicolumn{9}{c}{\MODEL \ -- PaliGemma Initialized} \\
\MODEL-3B$^\text{Gemma}$ & 3B & 64.9 & 53.5 & 70.9 & 79.5 & 68.6 & 61.3 & 65.4 \\
\midrule
\rowcolor{gray!15}
\multicolumn{9}{c}{\MODEL \ -- Llama-3.2-Vision Initialized} \\
\MODEL-11B & 11B & 66.4 & 42.1 & 74.3 & \underline{91.6} & 65.7 & 64.3 & 65.1 \\
\midrule
\rowcolor{gray!15}
\multicolumn{9}{c}{\MODEL \ -- Qwen2.5-VL Initialized} \\
\MODEL-3B & 3B & 62.7 & 68.1 & 71.9 & 79.6 & 73.5 & 63.8 & 69.1 \\
\MODEL-7B & 7B & \underline{71.3} & \underline{74.2} & \underline{78.7} & 85.4 & \underline{81.8} & \underline{70.0} & \underline{76.6} \\ 
\MODEL-32B & 32B & \textbf{73.7} & \textbf{78.6} & \textbf{78.9} & 88.1 & \textbf{82.8} & \textbf{73.7} & \textbf{78.7} \\

\bottomrule
\end{tabular}
\vspace{-0.2cm}
\label{tab:mmeb}
\end{table}

\subsection{Main Results}
\label{sec:main_results}

We report the overall multimodal embedding performance of different \MODEL variants and baseline methods on MMEB in Table~\ref{tab:mmeb}. 
Similarly, we present the visual document retrieval performance of \MODEL and baselines on ViDoRe v2 in Table~\ref{tab:vidore-v2}. 
All \MODEL results are reported with 16 query-side vectors and 64 candidate-side vectors, \ie $(r_q, r_c) = (16,64)$, and we will discuss the impact of the number of \EMBED used in \S\ref{sec:ablations}.
We conclude key observations from those metrics as follows.

\custompara{\MODEL delivers substantial improvements over the best existing single-vector baselines at comparable model sizes.} At the 3B scale, \MODEL achieves 69.1 overall on MMEB, already surpassing MoCa-3B (67.5) with +1.6\% relative improvement. 
At 7B, the margin widens: \MODEL reaches 76.6, outperforming MoCa-7B (71.5) and mmE5 (69.8) by over 5-7 points. Scaling further to 32B yields 78.7 overall, a clear improvement over both the strongest baselines and our smaller variants.
Importantly, the relative gains of \MODEL increase with model size -- while the 3B variant offers competitive results, the 7B and 32B models establish new state-of-the-art performance with the gap over baselines widening as scale increases. This trend suggests that \MODEL scales more favorably than prior approaches, with benefits compounding in larger regimes.

\begin{table}[t]
\centering
\small
\setlength{\tabcolsep}{3.2pt}
\caption{NDCG@5 (\%) results on the ViDoRe v2 benchmark, which covers 7 tasks on visual document retrieval. 
``Syn'' denotes synthetic data, ``Mul'' indicates multilingual tasks, and ``Bio'' refers to biomedical domains.
}
\vspace{3pt}
\begin{tabular}{lccccccccc}
\toprule
Models & Size & ESG\_Human & Eco\_Mul & Bio & ESG\_Syn & ESG\_Syn\_Mul & Bio\_Mul & Eco & Avg. \\
\midrule
\rowcolor{gray!15}
\multicolumn{10}{c}{Single-Vector Retrieval} \\
SigLIP        & 652M & 28.8 & 14.0 & 33.8 & 19.8 & 21.9 & 18.2 & 29.8 & 23.8 \\
VLM2Vec         & 7B & 33.9 & 42.0 & 38.8 & 36.7 & 38.4 & 29.7 & 51.4 & 38.7 \\
VisRAG-Ret         & 3B & 53.7 & 48.7 & 54.8 & 45.9 & 46.4 & 47.7 & 59.6 & 51.0 \\
GME              & 7B & \textbf{65.8} & 56.2 & \underline{64.0} & 54.3 & 56.7 & 55.1 & \underline{62.9} & 59.3 \\
mmE5            & 11B & 52.8 & 44.3 & 51.3 & 55.1 & 54.7 & 46.8 & 48.6 & 50.5 \\
MoCa-3B & 3B & 63.3 & \underline{57.3} & 62.5 & 58.3 & 54.8 & 59.8 & 62.8 & 59.8 \\
MoCa-7B & 7B & 58.8 & \textbf{57.6} & 63.2 & 55.3 & 51.4 & \underline{61.3} & \textbf{63.8} & 58.8 \\
\midrule
\rowcolor{gray!15}
\multicolumn{10}{c}{Multi-Vector Retrieval} \\
ColPali & 3B & 51.1 & 49.9 & 59.7 & 57.0 & 55.7 & 56.5 & 51.6 & 54.5 \\
ColQwen2 & 2B & 62.2 & 53.2 & 61.8 & 53.4 & 54.2 & 56.5 & 61.5 & 57.5 \\
\midrule
\rowcolor{gray!15}
\multicolumn{10}{c}{\MODEL} \\
\MODEL-3B & 3B & 63.7 & 55.5 & 61.7 & \underline{62.6} & \underline{57.4} & 58.7 & 62.3 & \underline{60.3} \\
\MODEL-7B & 7B & \underline{62.9} & 54.2 & \textbf{65.0} & \textbf{62.9} & \textbf{61.1} & \textbf{61.9} & 60.9 & \textbf{61.3} \\
\bottomrule
\end{tabular}
\vspace{-0.3cm}
\label{tab:vidore-v2}
\end{table}

\custompara{The choice of VLM backbone has a pronounced effect on \MODEL performance across tasks.}
\MODEL-11B with Llama-3.2-Vision backbone shows strong grounding and solid classification abilities, but its VQA score drops sharply to 42.1 -- more than 32 points lower than the Qwen2.5-VL-initialized 7B model (74.2). This limitation caps its overall score at 65.1 despite excelling in other subtasks. In contrast, Qwen2.5-VL initialization consistently delivers balanced improvements across all metrics: \MODEL-7B achieves 76.6, and scaling further to 32B pushes the state of the art to 78.7, with especially strong VQA and retrieval capabilities.
We notice that if the underlying base model itself struggles on some domains when used as a generative model, such a weakness directly propagates into \MODEL as an embedding model. 
For example, \cite{huang2025mimicking} suggests that Llama-3.2-Vision-11B is less competitive in most zero-shot VQA benchmarks, and such weakness is inherited in \MODEL-11B.

\custompara{\MODEL demonstrates strong retrieval performance on ViDoRe v2, particularly in multilingual and biomedical domains, despite not being trained on multilingual data.}
Even at the 3B scale, \MODEL matches or surpasses much larger baselines, showing robustness across all seven evaluation tracks. When scaled to 7B, the model yields further gains, with the largest improvements appearing in multilingual and biomedical domains. This is especially noteworthy given that no explicit multilingual data was included during training, suggesting that \MODEL effectively retains and leverages cross-lingual capabilities from its backbone. 

\begin{figure*}[t]
    \begin{center}
        \includegraphics[width=\linewidth]{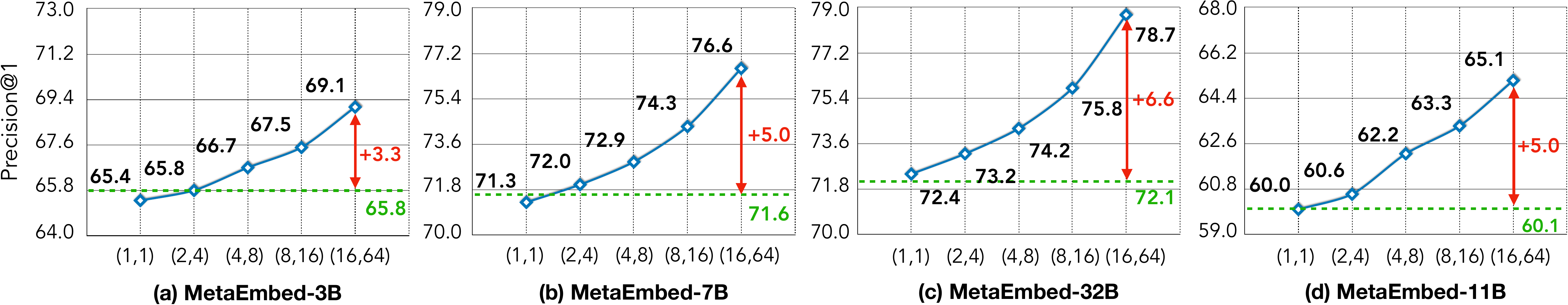}
    \end{center}
    \vspace{-5pt}
    \caption{Impact of retrieval budget on MMEB across \MODEL of varying model sizes. Retrieval budget is denoted as $(r_q, r_c)$, \ie a tuple of the number of \EMBED used on query and candidate side. Increasing the retrieval budget from (1,1) to (16,64) consistently improves performance for all model sizes, with larger gains observed in higher-capacity models. The dashed green lines indicate the best single-vector retrieval performance and red arrows indicate the absolute gain (in percentage points) between \MODEL and single-vector retrieval.}
    \label{fig:testing_scale}
    \vspace{-1em}
\end{figure*}

\subsection{Ablation Studies}
\label{sec:ablations}

To better understand \MODEL, we design comprehensive ablation studies to investigate its test-time scaling capabilities, the effectiveness of MMR and its robustness across different models.

\begin{wrapfigure}{l}{4.5cm}
  \centering
  \vspace{-1em}
  \begin{subfigure}{\linewidth}
    \centering
    \includegraphics[width=\linewidth]{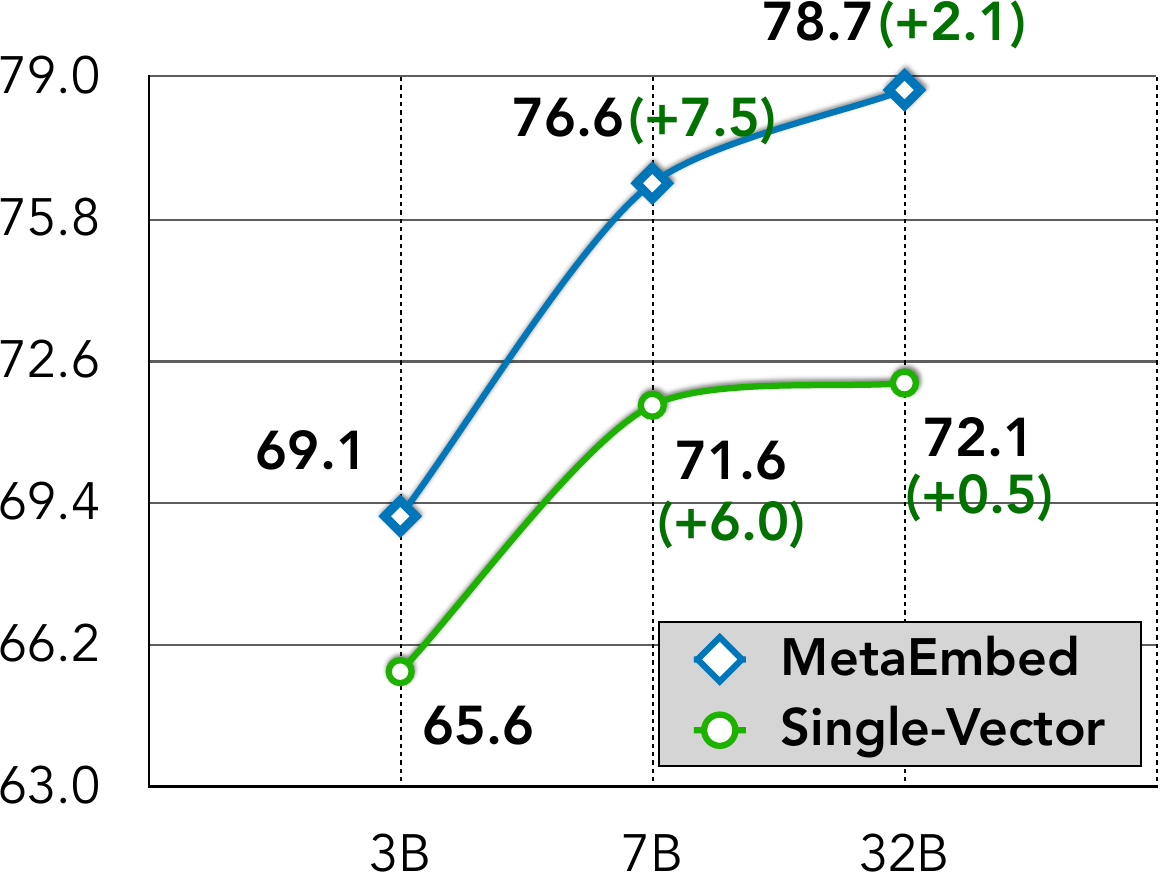}
    \caption{\MODEL with $(16,64)$ retrieval budget shows less diminishing returns as model size scales. Green numbers denote the gain compared to the preceding model size. }
    \label{fig:training_scale}
  \end{subfigure}
  
  \vspace{1.5em}
  \begin{subfigure}{\linewidth}
    \centering
    \includegraphics[width=\linewidth]{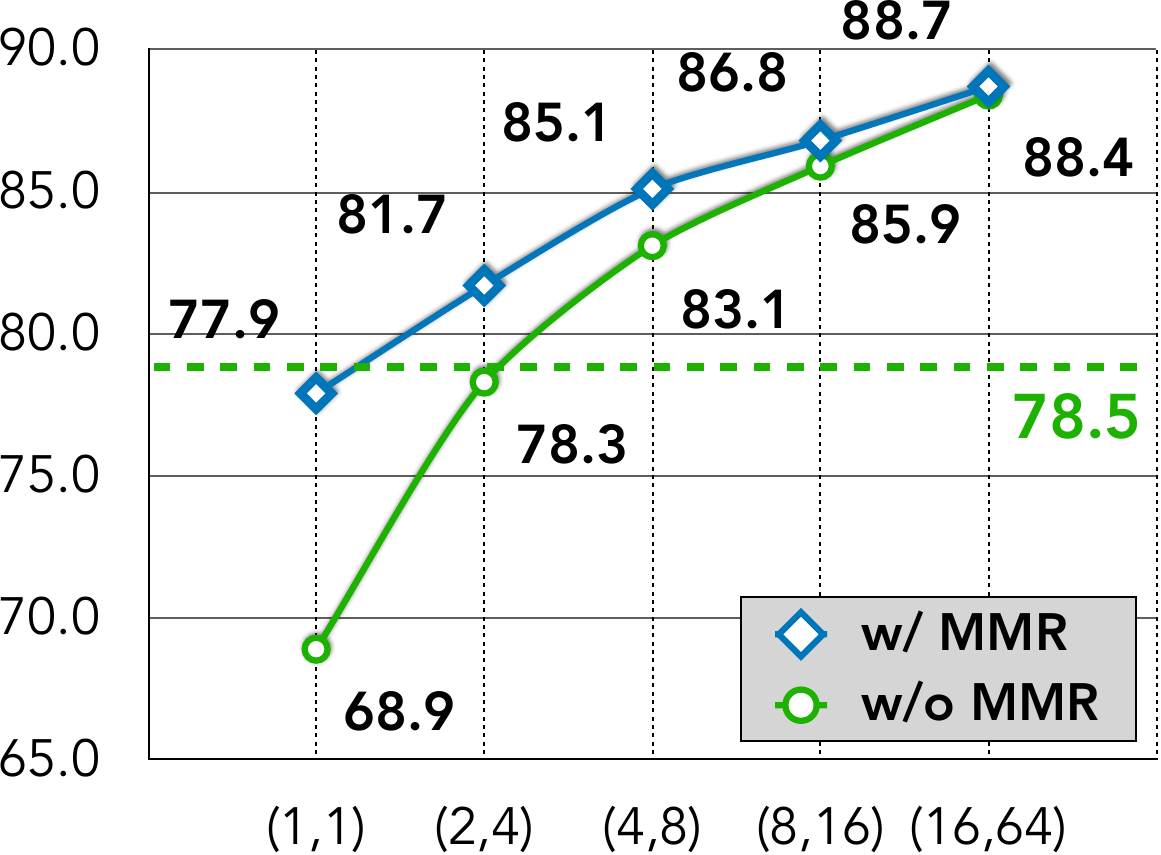}
    \caption{Average NDCG@5 (\%) on ViDoRe v1 benchmark with varying retrieval budgets on \MODEL-3B with and without MMR design.}
    \label{fig:mmr_compare}
  \end{subfigure}
  \caption{Ablation studies.}
  \vspace{-2em}
\end{wrapfigure}

\newpara{How does the performance scale with the retrieval cost? }
We present the performance plots of different \MODEL models with respect to retrieval cost in Fig.~\ref{fig:testing_scale}. 
Data points in each plot correspond to \MODEL performance with a specific model size with varying test-time retrieval budgets. The dashed green line marks the best-performance single-vector retrieval model with identical training settings. 
The plots show that across model sizes, the curves rise steadily as more retrieval budget is allocated.
While the improvement is modest for smaller models as \MODEL-3B shows +3.3 points relative gain against the single-vector method, we observe that it becomes more noticeable as the base model size grows and \MODEL brings the most pronounced improvements on the 32B model with a +6.6 points gain. 

\newpara{Does \MODEL apply to pre-trained VLMs of different sizes and architectures? }
Fig.~\ref{fig:testing_scale} (d) already demonstrates that \MODEL-11B shows advantages when finetuned on different VLM, Llama-3.2-Vision-11B, showing the robustness of our method across architectures.
Another key observation is that \MODEL makes more effective use of larger model capacity compared to single-vector methods. Fig.~\ref{fig:training_scale} presents the performance of the two approaches on MMEB under identical training settings as model size increases. We find that \MODEL achieves more substantial gains than single-vector retrieval. Notably, the improvement of the single-vector baseline from 7B to 32B is no longer statistically significant while \MODEL still holds a noticeable gain.

\newpara{How effective is the MMR design? }
To investigate how MMR functions, \ie how it organizes query and candidate information in a nested order of importance, we use average NDCG@5 on ViDoRe-v1 and report test-time scaling curves on two variants of \MODEL-3B: with and without MMR in Fig.~\ref{fig:mmr_compare}. 
If MMR is not enabled during training, the flexibility of \MODEL will be severely constrained, as evidenced by the substantial performance drop under low retrieval budgets. 
For example, the performance drop hits 9.0 points when using \MODEL without MMR as a single-vector retrieval model, \ie $(r_q, r_c)=(1,1)$. 
Although the gap narrows as retrieval budget increases, the model with MMR consistently outperforms the non-MMR model as shown in the figure.
Surprisingly, we find that even at the full budget $(r_q, r_c)=(16,64)$, the MMR model still performs slightly better, demonstrating that MMR does not sacrifice the original multi-vector retrieval ability at full scale.
We additionally report the test-time scaling results on MMEB in Appendix~\ref{app:detailed_ablation_results} where MMR shows negligible performance degradation. 

\section{Discussion}
\label{sec:discussion}

In this section, we mainly discuss the efficiency of \MODEL as a flexible multi-vector retrieval method, with a focus on index memory consumption and latency under varying retrieval budgets.

A typical online retrieval process consists of three stages: 
(a) Query encoding, where the query is processed by the encoder to obtain contextualized embeddings. 
(b) Scoring, where query embeddings are compared with candidate document embeddings in the index. For single-vector dense retrieval, this is a dot-product operation between pairs of vectors, while in multi-vector retrieval such as \MODEL it requires late interaction (e.g., MaxSim in Eq.~\ref{eq:late_interaction}) between multiple embeddings. (c) Ranking, a lightweight operation where candidate documents are sorted based on the scores.

We report the efficiency analysis of \MODEL-7B in Table~\ref{tab:retrieval_latency} with the following observations: 
\begin{enumerate}[leftmargin=*]
    \item Although the number of scoring FLOPs grows substantially with larger retrieval budgets, the scoring stage itself is not compute-bounded until the extreme case of $(16,64)$. The measured latencies remain nearly flat across moderate budgets, demonstrating that GPU throughput can accommodate the additional FLOPs without becoming a bottleneck.
    \item The relative contribution of scoring cost to the overall retrieval pipeline is negligible compared to query encoding. For instance, encoding an image query of 1024 tokens requires 42.72 TFLOPs and 788ms. These figures are orders of magnitude larger than the scoring costs reported in Table~\ref{tab:retrieval_latency}, indicating that efficiency improvements should primarily target encoding rather than scoring with small number of candidates.
    \item As a flexible multi-vector retrieval method, index memory consumption can grow proportionally with the retrieval budget. While this can present challenges for large deployments, the issue can be mitigated by using a balanced retrieval budget or more frequent offloading of index data to CPU memory.
\end{enumerate}

Overall, these findings suggest that \MODEL is efficient in practice. Query encoding dominates latency, scoring is lightweight under most realistic budgets with a small number of candidates, and memory scaling can be controlled by either selecting balanced retrieval budgets or system-level strategies such as CPU swapping.

\begin{table}[t]
\centering
\setlength{\tabcolsep}{2.7pt}
\caption{Efficiency analysis of \MODEL-7B with different retrieval budgets on an A100 GPU with 100,000 candidates per query with scoring batch size of 1,000. 
Query encoding and index generation latency are omitted because they remain the same for all variants. Latency refers specifically to scoring latency and mean and standard deviation of latency are reported with 10 runs. Index is stored and compared with \texttt{bfloat16} precision~\citep{wang_kanwar_2019_bfloat16}.}
\label{tab:retrieval_latency}
\begin{tabular}{lcccc}
\toprule
\textbf{Retrieval Budget} & \textbf{Scoring FLOPs (G)} & \textbf{Latency (ms)} & \textbf{Index Memory (GiB)} & \textbf{MMEB Acc (\%)} \\
\midrule
$(1,1)$ & 0.71 & 1.67${_{\pm\text{0.13}}}$ & 0.68 & 71.3 \\
$(2,4)$ & 5.73 & 1.66${_{\pm\text{0.12}}}$ & 2.67 & 72.0 \\
$(4,8)$ & 22.94 & 1.67${_{\pm\text{0.12}}}$ & 5.34 & 72.9  \\
$(8,16)$ & 91.75 & 1.92${_{\pm\text{0.12}}}$ & 10.68 &  74.3 \\
$(16,64)$ & 733.89 & 6.25${_{\pm\text{0.07}}}$ & 42.72 &  76.6 \\
\bottomrule
\end{tabular}
\vspace{-1em}
\end{table}

\section{Conclusion}
We present \MODEL, a new paradigm for multimodal retrieval that rethinks the construction and interaction of embeddings at scale. By leveraging a small set of learnable \TOKEN and training them through our proposed Matryoshka Multi-Vector Retrieval (MMR) framework, \MODEL organizes information into coarse-to-fine levels of granularity. This design enables flexible late interaction that balances retrieval accuracy, index size, and latency -- unlocking test-time scalability for multimodal retrieval.
We believe \MODEL opens a path toward more general, efficient, and controllable multimodal retrieval, bridging the gap between fine-grained expressiveness and large-scale deployability.

\vspace{1em}
\noindent{\bf Acknowledgements. }
We thank Anshumali Shrivastava, Xueyuan Su, Xu Han and Norman Huang for insightful discussions and support.

\clearpage
\newpage
\bibliographystyle{assets/plainnat}
\bibliography{paper}

\clearpage
\newpage
\beginappendix

\section{Implementation Details}
\label{app:imple}

\begin{table}[h]
\centering
\caption{Training details of \MODEL variants.}
\label{tab:impl_details}
\setlength{\tabcolsep}{5pt}
\begin{tabular}{lccccc}
\toprule
\textbf{} & \textbf{Batch Size} & \textbf{Learning Rate} & \textbf{Training Cost} & \textbf{Embedding Dim} \\
\midrule
\MODEL-3B$^\text{Gemma}$ & 2,048 & $1 \times 10^{-4}$ & 32 H100s for 14h & 2,048 \\
\MODEL-3B & 2,048 & $1 \times 10^{-4}$ & 32 H100s for 23h & 2,048 \\
\MODEL-7B & 1,536 & $1 \times 10^{-4}$ & 32 H100s for 30h & 3,584 \\
\MODEL-11B & 1,024 & $1 \times 10^{-4}$ & 32 H100s for 10h & 4,096 \\
\MODEL-32B & 1,536 & $1 \times 10^{-5}$ & 64 H100s for 25h & 5,120 \\
\bottomrule
\end{tabular}
\end{table}

We list the details of each \MODEL variant in Table~\ref{tab:impl_details}.
We reserve 1\% training data from each subset of MMEB-train as evaluation split and training was early stopped when evaluation loss stops dropping. 
All models are trained with gradient checkpointing~\citep{chen2016training} to reduce memory usage. 
Training was conducted using PyTorch~\citep{paszke2019pytorch} \texttt{2.6.0+cu124} and FlashAttention 2.0~\citep{dao2024flashattention}.
For the 3B configuration, we adopt Distributed Data Parallel~\citep{li2020pytorch} while for all other model sizes we use Fully Sharded Data Parallel (FSDP)~\citep{zhao2023pytorch} v2.
To prevent distributed hanging when training samples contain no images under FSDP, we pad those samples with a placeholder image to ensure the visual encoder is activated on each GPU.

\section{Detailed MMEB Ablation Results}
\label{app:detailed_ablation_results}

\begin{table}[h]
\centering
\setlength{\tabcolsep}{5pt}
\caption{Comparison between \MODEL and single-vector \& multi-vector retrieval models trained with identical settings. NoMMR indicates Matryoshka Multi-Vector Retrieval (MMR) is disabled. $\Delta$ denotes the difference to the best single-vector retrieval method, \ie single-last.}
\begin{tabular}{lccc|ccc|ccc|ccc}
\toprule
 & \multicolumn{3}{c}{3B} & \multicolumn{3}{c}{7B} & \multicolumn{3}{c}{32B} & \multicolumn{3}{c}{11B} \\
\cmidrule(lr){2-4}\cmidrule(lr){5-7}\cmidrule(lr){8-10}\cmidrule(lr){11-13}
Type & MMEB & $\Delta$ &  & MMEB & $\Delta$ &  & MMEB & $\Delta$ &  & MMEB & $\Delta$ &  \\
\midrule
single-last & 65.6 & 0   & & 71.6 & 0   & & 72.1 & 0   & & 60.1 & 0   & \\
single-mean & 65.2 & -0.4 & & 71.2 & -0.4 & & 71.1 & -1.0 & & 58.1 & -2.0 & \\
split-$(16,64)$ & 64.2 & -1.4 & & 70.1 & -1.5 & & 70.5 & -1.6 & & 56.0 & -4.1 & \\
\midrule
\rowcolor{gray!15}
\multicolumn{13}{c}{\MODEL} \\
$(1,1)$   & 65.4 & -0.2 & & 71.3 & -0.3 & & 72.4 & +0.3 & & 60.0 & -0.1 & \\
$(2,4)$   & 65.8 & +0.2 & & 72.0 & +0.4 & & 73.2 & +1.1 & & 60.6 & +0.5 & \\
$(4,8)$   & 66.7 & +1.1 & & 72.9 & +1.3 & & 74.2 & +2.1 & & 62.2 & +2.1 & \\
$(8,16)$   & 67.5 & +1.9 & & 74.3 & +2.7 & & 75.8 & +3.7 & & 63.3 & +3.2 & \\
$(16,64)$   & 69.1 & +3.5 & & 76.6 & +5.0 & & 78.7 & +6.6 & & 65.1 & +5.0 & \\
\midrule
NoMMR-$(16,64)$  & 69.3 & +3.7 & & 77.0 & +5.4   & & 79.1 & +7.0   & & 66.2 & +6.1   & \\
\bottomrule
\end{tabular}
\end{table}

To further compare the superiority of \MODEL against both single-vector and multi-vector retrieval methods, we design the following baselines based on the same pre-trained models for fair comparison:  
\begin{enumerate}[leftmargin=*]
    \item single-last: a single-vector retrieval method that uses the last-token hidden state from the last layer as the retrieval representation. 
    \item single-mean: similar to the above, but applies average pooling over all last-layer hidden states to obtain the retrieval representation. 
    \item split-$(16,64)$: a simple multi-vector retrieval baseline where the query-side last-layer hidden states are evenly partitioned into 16 segments, with the mean of each segment taken as 16 query vectors; the same process produces 64 candidate-side vectors. This method does not introduce additional parameters, making it a suitable fixed-length multi-vector retrieval baseline. 
\end{enumerate}

Our findings indicate that \MODEL goes beyond test-time scaling advantages mentioned in the main text: it consistently outperforms the top single-vector method as well as a naive multi-vector baseline. Moreover, the MMR design brings no statistically significant loss, demonstrating that it adds flexibility while maintaining retrieval quality.

\section{Baseline Method Introduction}
\label{app:baseline_method_intro}

For clarity and completeness, we provide short introductions to the baseline methods considered in Table~\ref{tab:mmeb}. 
All performance metrics reported on baseline methods are directly taken from the corresponding original papers or \cite{chen2025moca}.

\paragraph{CLIP~\citep{CLIP}.}  
CLIP is a dual-encoder trained with contrastive learning on 400M image--text pairs.  
It learns aligned representations for both modalities, enabling strong zero-shot classification and retrieval capabilities.

\paragraph{MagicLens~\citep{MagicLens}.}
MagicLens is a lightweight dual-encoder for instruction-guided image retrieval. It is trained in a self-supervised manner on roughly 36.7M (query-image, text instruction, target-image) triplets mined from co-occurring web images. 

\paragraph{UniIR~\citep{UniIR}.}
UniIR is a unified, instruction-guided multimodal retriever that handles eight retrieval task formats spanning text, image, and mixed-modality queries/candidates. It is jointly trained on ten heterogeneous datasets, showing robust in-distribution performance and zero-shot generalization across tasks.

\paragraph{MM-Embed~\citep{mmembed}.}  
MM-Embed converts a multimodal large language model into a universal bi-encoder for retrieval.  
It is fine-tuned with modality-aware hard negatives across diverse retrieval datasets to improve cross-modal alignment.

\paragraph{GME~\citep{zhang2024gme}.}  
The General Multimodal Embedder is trained on a large synthetic dataset containing diverse multimodal queries and documents.  
It introduces fused-modal training examples (mixed text--image inputs) to enable universal any-to-any modality retrieval.

\paragraph{VLM2Vec~\citep{jiang2025vlmvec}.}  
VLM2Vec transforms a pretrained vision-language model into a universal embedding model through instruction-tuned contrastive learning.  
It is trained on the Massive Multimodal Embedding Benchmark (MMEB), covering 36 tasks across classification, VQA, retrieval, and grounding.

\paragraph{MMRet~\citep{megapairs}.}  
MMRet builds on the MM-Embed framework but introduces further refinements in negative sampling and large-scale fine-tuning on massive synthetic dataset, making it one of the strongest retrieval-centric embedding models in prior work.

\paragraph{mmE5~\citep{mme5}.}  
mmE5 extends the multilingual text embedding model E5 into the multimodal setting. 
It is trained with multilingual and multimodal signals, leveraging synthetic image-text pairs and hard negatives, and achieves strong state-of-the-art results on MMEB prior to more recent models.

\paragraph{MoCa~\citep{chen2025moca}.}  
MoCa (Modality-aware Causal Pre-training) introduces a two-stage process: modality-aware continual pre-training to adapt causal vision--language models for bidirectional encoding, followed by heterogeneous contrastive fine-tuning across text, image, and mixed modality pairs.  
We evaluate both MoCa-3B and MoCa-7B, which show competitive overall performance among baselines on MMEB.

\paragraph{B3~\citep{B3}}  
B3-7B is a 7B-parameter instruction-tuned multimodal retriever with specialized batch mining techniques. It achieves strong results on MMEB and serves as an additional competitive baseline.

\end{document}